\documentclass[aps,prd,twocolumn,showpacs,groupedaddress,nofootinbib]{revtex4}
\usepackage{graphicx}

\begin{document}


\title{Phase structures of strong coupling lattice QCD with overlap fermions at finite temperature and chemical potential}

\author{Xiao-Lu Yu}
\address{Department of Physics, Zhongshan (Sun Yat-Sen) University,
Guangzhou 510275, China}

\author{Xiang-Qian Luo}
\thanks{Corresponding author. Email address: stslxq@zsu.edu.cn}
\address{
CCAST (World Laboratory), P.O. Box 8730,
Beijing 100080, China\\
Department of Physics, Zhongshan (Sun Yat-Sen) University,
Guangzhou 510275, China}
\thanks{Mailing address.}

\date{\today}

\begin{abstract}
We perform the first study of lattice QCD with overlap fermions at
finite temperature $T$ and chemical potential $\mu$. We start from
the Taylor expanded overlap fermion action, and derive in the
strong coupling limit the effective free energy by mean field
approximation. On the ($\mu,T$) plane and in the chiral limit,
there is a tricritical point, separating the second order chiral
phase transition line at small $\mu$ and large $T$, and first
order chiral phase transition line at large $\mu$ and small $T$.
\end{abstract}



\maketitle

\section{Introduction}

To study the nature of matter under extreme conditions in QCD is
one of the most challenging issues in particle physics. Several
novel phases have been suggested, such as quark-gluon plasma (QCP)
\cite{Back:2003ns} and color superconductivity\cite{Rajagopal}.
Precise determination of the QCD phase diagram on the ($\mu,T$)
plane will provide valuable information for these novel phases.

Numerical simulation of lattice gauge theory (LGT) is the most
reliable nonperturbative method based on the first principles.
This approach has been successfully applied to the findings of the
chiral and deconfinement phase transitions\cite{Meyer} at finite
$T$ with zero $\mu$. However, LGT experiences serious problems,
like species doubling with naive fermions and complex action at
finite $\mu$.

The Hamiltonian formulation of LGT at finite density
\cite{Gregory:1999pm,Luo:2000xi,Fang:2002rk,Luo:2004mc} does not
have the complex action problem. The complex action problem in
Lagrangian formulation forbids numerical simulation at real $\mu$.
The recent years have seen enormous
efforts\cite{Fodor:2002hs,deForcrand:2002ci,D'Elia:2002gd,Azcoiti:2005ua,Chen:2004tb}
on solving the complex action problem, and some very interesting
information on the phase diagram for QCD at large $T$ and small
$\mu$ has been obtained. QCD at large $\mu$ is of particular
importance for neutron star or quark star physics.

There have been several popular approaches to solving the species
doubling problem of naive fermions. The staggered fermion approach
preserves the remnant of chiral symmetry, but it breaks the flavor
symmetry and doesn't completely solve the species doubling
problem. The Wilson fermion approach avoids the doublers and
preserve the flavor symmetry, but it explicitly breaks the chiral
symmetry; In order to define the chiral limit, one has to do
nonperturbative fine-tuning of the bare fermion mass, which seems
to be an unnatural method from the physical point of
view\cite{Chandrasekharan}.

These years have seen increasing interest in the overlap fermion
approach\cite{Narayanan,Neuberger}, which is claimed to have the
properties that chiral symmetry is preserved and species doubling
problem may be solved\cite{luscher}.  Chiral symmetry in this
approach is not the original form, and therefore one of the
conditions of the Nielsen-Ninomiya (NO-GO) theorem\cite{Nielsen}
is not satisfied. However, the Dirac operator of overlap fermion
is nonlocal, and the computational costs\cite{Fodor:2003bh} for
simulating dynamical overlap fermions are typically two orders of
magnitude heavier than for the Wilson or KS formulations. It is
also very tough to introduce the chemical potential into the
action. Before the breakthrough of numerical algorithms for
applying overlap fermions to QCD thermodynamics, it is very useful
to do an analytical study. At $\mu=0$ and $T=0$, this was made
possible in Ref. \cite{Ichinose}, where the generalized overlap
fermion approach was introduced by Taylor expansion of the Dirac
operator.

In this paper, we study the phase diagram of the strong coupling
lattice QCD on the ($\mu,T$) plane using Taylor expanded overlap
fermions. In the chiral limit, we elucidate the phase structure
and find lines of first and second order chiral phase transitions,
as well as the tricritical point where the two lines join.

The rest of paper is organized as follows. In Sec. \ref{sec2}, we
shall give a brief review of the Taylor expanded overlap fermion
approach, and rewrite the fermion action in terms of composite
operators which transform covariantly under the extended chiral
transformation. In Sec. \ref{sec3}, we introduce temperature and
chemical potential into the system and derive the effective free
energy under mean field approximation. In Sec. \ref{sec4}, we
analyze the QCD phase diagram on the ($\mu,T$) plane in the chiral
limit. In Sec. \ref{sec5}, the results are summarized.

\section{Formulation of GO fermions}
\label{sec2}

The action of lattice QCD is given by $S=S_{G}+S_{F,m}$, where
\begin{eqnarray}
S_{G}&=& -{\beta \over 6} \sum_p {\rm Tr} (U_{p}
+U_{p}^{\dagger}-2),
\nonumber \\
S_{F,m}&=&S_{F}+m\sum{\bar\psi}(x)\psi(x).
 \label{eqn1}
\end{eqnarray}
Here $\beta=6/g^2$, and  $U_p$ is the ordered product of link
variables $U$ around an elementary plaquette.  $S_{F}$ is the
massless fermion action:
\begin{eqnarray}
S_{F}=a^{4}\sum_{x,y}{\bar\psi}(x)D(x,y)\psi(y), \label{eqn2}
\end{eqnarray}
with the overlap Dirac operator $D(x,y)$ defined
as\cite{Neuberger}
\begin{eqnarray}
aD=1+X\frac{1}{\sqrt{X^{\dag}X}}.
 \label{eqn3}
\end{eqnarray}
This operator is nonlocal and it is extremely difficult to do
analytical calculations. An alternative is proposed in Refs.
\cite{Ichinose,Nagao}, where the operator $X$ is expressed as
\begin{eqnarray}
X=A+B\sum_j \gamma_j \Gamma_{j}^{-} - C\sum_j \gamma_j
\Gamma_{j}^{+},
 \label{eqn4}
\end{eqnarray}
with
\begin{eqnarray}
\Gamma_{j}^{-}(x,y)&=&\delta_{x+j,y}U_{j}(x)-\delta_{x,y+j}U_{j}^{\dag}(y),
\nonumber \\
\Gamma_{j}^{+}(x,y)&=&\delta_{x+j,y}U_{j}(x)+\delta_{x,y+j}U_{j}^{\dag}(y),
\nonumber \\
A&=&\frac{1}{a}(4r-M_{0}),
\nonumber \\
B&=&\frac{rt}{2a},
\nonumber \\
C&=&\frac{t}{2a}.
 \label{eqn5}
\end{eqnarray}
Then, one may expand $D(x,y)$ in powers of the parameter $t$
as\cite{Ichinose,Nagao}
\begin{eqnarray}
aD_{x,y}=2\theta(A)\delta_{xy}+\frac{C} {\vert A
\vert}\sum\gamma_{j}\Gamma_{j}^{-}(x,y)+O(t^{2}),
 \label{eqn6}
\end{eqnarray}
where we keep the contributions up to $O(t^{2})$. In later discussions, we consider the
case\cite{Ichinose} of negative $A$, which is expected to have the desired properties of QCD.

In Ref. \cite{luscher}, it was shown that the Ginsparg-Wilson (GW)
relation\cite{Ginsparg}
\begin{eqnarray}
\gamma_{5}D+D\gamma_{5}=aD\gamma_{5}D
 \label{eqn7}
\end{eqnarray}
implies an exact symmetry on the lattice, since the
Nielsen-Ninomiya theorem is not applicable. It has been verified
\cite{Ichinose,Nagao} that the $t$-expanded $D(x,y)$ given by Eq.
(\ref{eqn6}) satisfies the GW relation.

The action is invariant under infinitesimal chiral transformation
$\psi \to \psi+\delta \psi$ and ${\bar \psi} \to {\bar
\psi}+\delta {\bar \psi}$, where \cite{luscher}
\begin{eqnarray}
\delta\psi=\varepsilon\gamma_{5}(1-aD)\psi, ~~
\delta{\bar\psi}=\varepsilon{\bar\psi}\gamma_{5}.
 \label{eqn8}
\end{eqnarray}
It is easy to show that
\begin{eqnarray}
\langle {\bar\psi}\gamma_{5}\psi \rangle  \to \langle
{\bar\psi}\gamma_{5}\psi \rangle +2\varepsilon\langle
{\bar\psi}\left(1-\frac{1}{2}aD\right)\psi \rangle   \label{eqn9}
\end{eqnarray}
under the transformation (\ref{eqn8}).
 This means that for $m=0$, the vacuum has an exact chiral symmetry
 if the vacuum expectation value
\begin{eqnarray}
\langle {\bar\psi}\left(1-\frac{1}{2}aD\right)\psi \rangle =0 .
 \label{eqn10}
\end{eqnarray}
Otherwise, the chiral symmetry is spontaneously broken if
\begin{eqnarray}
\langle {\bar\psi}\left(1-\frac{1}{2}aD\right)\psi \rangle \neq 0.
\label{eqn11}
\end{eqnarray}
These properties are very important for studying the chiral phase
transition at finite temperature and chemical potential.

Therefore, we could choose the quantity $\langle
{\bar\psi}(1-aD/2)\psi \rangle $ as the chiral order parameter,
which will reduce to the conventional one in the continuum limit
$a \to 0$. It is more convenient to use the fermion fields $q$ and
${\bar q}$. They are related to $\psi$ and ${\bar \psi}$ by
\cite{Nagao}
\begin{eqnarray}
{\bar  q}={\bar \psi},  ~~ q=\left(1-\frac{1}{2}aD\right)\psi .
\label{eqn12}
\end{eqnarray}
The extended chiral transformation is defined by
\begin{eqnarray}
\delta q=\varepsilon\gamma_{5}q, ~~ \delta{\bar
q}=\varepsilon{\bar q}\gamma_{5}.
 \label{eqn13}
\end{eqnarray}
The chiral order parameter is then given by
\begin{eqnarray}
\langle {\bar  q}q \rangle =\langle {\bar
\psi}\left(1-\frac{1}{2}aD\right)\psi \rangle  . \label{eqn14}
\end{eqnarray}
For convenience, we set the lattice spacing $a=1$. Substituting
Eqs. (\ref{eqn6}), (\ref{eqn12}), and (\ref{eqn14}) into Eq.
(\ref{eqn1}), we could rewrite the fermion action in terms of the
new fermion fields ${\bar q}$ and $q$:
\begin{eqnarray}
S_{F,m}&=&\left(1+\frac{m}{2}\right)\sum_{x,y}{\bar
q}(x)D(x,y)q(y)
\nonumber\\
 &+&m\sum_{x}{\bar  q}(x)q(x)+O(t^{2}) .
 \label{eqn15}
\end{eqnarray}

\section{Effective free energy of GO fermion}
\label{sec3}

In the strong coupling limit, the gluonic action $S_G$ vanishes,
and $S \to S_{F,m}$. We introduce the chemical potential $\mu$
into the action Eq. (\ref{eqn15}) by replacing the link variables
in the temporal direction\cite{Hasenfratz} with $U_4(x)\rightarrow
e^{\mu} U_4(x)$, and $U^{\dag}_4(x)\rightarrow e^{-\mu}
U^{\dag}_4(x)$:
\begin{eqnarray}
S_{F,m}&=&\left(1+\frac{m}{2}\right)\frac{C}{\vert A\vert}
\bigg(\sum_{x}\sum_{j=1}^{d}[{\bar q}(x)\gamma_{j}U_{j}(x)
q(x+\widehat{j})
\nonumber\\
 & & ~~~~~~~~~~~~~~~~~~~~~~~~ -{\bar  q}(x+\widehat{j})\gamma_{j}U_{j}^{\dag}(x)q(x)]
 \nonumber\\
 & & ~~~~~~~~~~~~~~~~~+\sum_{x}[e^{\mu}{\bar q}(x)\gamma_{4}U_{4}(x)q(x+\widehat{4})
 \nonumber\\
  & & ~~~~~~~~~~~~~~~~~~~~~~~ -e^{-\mu}{\bar q}(x+\widehat{4})\gamma_{4}U_{4}^{\dag}(x)q(x)] \bigg)
\nonumber\\
 &+&m\sum_{x}{\bar q}(x)q(x).
 \label{eqn16}
\end{eqnarray}
Here $d=3$ is the spatial dimensions.

We use a notation $x=({\vec x},\tau)$ in which ${\vec x}$ and
$\tau$ represents the spatial and temporal coordinate
respectively. The temperature is given by $T=N_{\tau}^{-1}$ with
$N_{\tau}$ the number of temporal lattice sites. We also use a
representation of $\gamma$ matrices described in Ref.
\cite{Lurie}.

The partition function of the system is
\begin{eqnarray}
Z&=&\int D[U_{j}]D[U_{4}]D[{\bar  q}]D[q]\exp(-S_{F,m}) .
\label{eqn17}
\end{eqnarray}

To derive the effective free energy $F_{eff}=-\ln Z/\sum_{\vec
x}$, we first integrate over spatial gauge link variables $U_{j}$,
using the $SU(N_{c})$ group integration formulas and Taylor
expansion:
\begin{eqnarray}
&\int& D[U_{j}]
\exp\bigg(-\left(1+\frac{m}{2}\right)\frac{C}{\vert
A\vert}\sum_{x}\sum_{j=1}^{d}
\nonumber\\
&\times& [{\bar q}(x)\gamma_{j}U_{j}(x)q(x+\widehat{j})
- {\bar q}(x+\widehat{j})\gamma_{j}U_{j}^{\dag}(x)q(x)]\bigg)
\nonumber\\
& \approx & 1-\left(1+\frac{m}{2}\right)^{2}\left(\frac{C}{
A}\right)^{2}\frac{1}{N_{c}}
\nonumber\\
&\times& \sum_{x}\sum_{j=1}^{d}{\bar
q}(x)\gamma_{j}q(x+\widehat{j}){\bar
q}(x+\widehat{j})\gamma_{j}q(x),
\label{eqn18}
\end{eqnarray}
which should be a good approximation for $N_f/N_c < 3$.

Next, we linearize the four-fermion term in Eq. (\ref{eqn18}) by
mean field approximation\cite{Fang:2002rk}
\begin{eqnarray}
&&\sum_{x}\sum_{j=1}^{d}{\bar
q}(x)\gamma_{j}q(x+\widehat{j}){\bar
q}(x+\widehat{j})\gamma_{j}q(x)
\nonumber\\
&\sim&\sum_{x}d[N_{f}({\bar v})^{2}-2{\bar v}{\bar q}(x)q(x)],
 \label{eqn19}
\end{eqnarray}
where ${\bar v}$ stands for the chiral condensate $\langle {\bar
q} q \rangle $. Substituting Eqs. (\ref{eqn18}) and (\ref{eqn19})
into Eq. (\ref{eqn17}), the partition function becomes
\begin{eqnarray}
Z=\int D[U_{4}]D[{\bar  q}]
D[q]\exp\left[-S^{eff}_{F,m}\left(U_{4},{\bar q},q;{\bar
v}\right)\right]
 \label{eqn20}
\end{eqnarray}
with
\begin{eqnarray}
&&S^{eff}_{F,m}\left(U_{4},{\bar  q},q;{\bar v}\right)
\nonumber\\
&=&\left(1+\frac{m}{2}\right)\frac{C}{\vert A\vert}\
\sum_{x}[e^{\mu}{\bar q}(x)\gamma_{4}U_{4}q(x+\widehat{4})
\nonumber\\
&-& e^{-\mu}{\bar
q}(x+\widehat{4})\gamma_{4}U_{4}^{\dag}q(x)]
\nonumber\\
&+&\left(1+\frac{m}{2}\right)^{2}\frac{dC^{2}}{N_{c}A^{2}}\sum_{x}d[N_{f}({\bar
v})^{2}-2{\bar v}{\bar  q}(x)q(x)]
\nonumber\\
&+&m\sum_{x}{\bar  q}(x)q(x). \label{eqn21}
\end{eqnarray}
The hopping terms between ${\bar  q}$ and $q$ exist only in the
temporal direction.

To compute the remaining integrations, we make a Fourier
transformation of the fermion fields
\begin{eqnarray}
q({\vec
x},\tau)&=&\frac{1}{\sqrt{N_{\tau}}}\sum_{n=1}^{N_{\tau}}\exp(ik_{n}\tau)\widetilde{q}({\vec
x},n),
\nonumber\\
{\bar q}({\vec
x},\tau)&=&\frac{1}{\sqrt{N_{\tau}}}\sum_{n=1}^{N_{\tau}}\exp(-ik_{n}\tau)\widetilde{{\bar
q}}({\vec x},n),
 \label{eqn22}
\end{eqnarray}
with $k_{n}=2\pi(n-1/2)/N_{\tau}$, and adopt the Polyakov
gauge\cite{Damgaard}
\begin{eqnarray}
U_{4}({\vec x},\tau)&=&{\rm diag} \left[e^{i\theta_{1}({\vec x})},e^{i\theta_{2}({\vec
x})},...,e^{i\theta_{N_{c}}({\vec x})}\right],
\nonumber\\
\sum_{a=1}^{N_{c}}\theta_{a}({\vec x})&=&0. \label{eqn23}
\end{eqnarray}
I.e., $U_{4}({\vec x},\tau)$ is diagonal and independent of
$\tau$.

Substituting Eqs. (\ref{eqn22}) and (\ref{eqn23}) into Eq.
(\ref{eqn20}) and integrating out the Grassmann
variables\cite{Nishida}, we have
\begin{eqnarray}
\int D[{\bar  q}]D[q]&&\exp
\left(-\frac{N_{f}}{N_{\tau}}\sum_{x}{\bar q}(x)\Delta q(x)\right)
\nonumber\\
&&=(\det\Delta)^{N_{f}/N_{\tau}},
 \label{eqn24}
\end{eqnarray}
where the determinant is
\begin{eqnarray}
\det\Delta&=&\prod_{{\vec
x}}\prod_{\alpha=1}^{N_{c}}\prod_{n=1}^{N_{\tau}}\bigg(\left[2\left(1+\frac{m}{2}\right)\frac{d}{N_{c}}
\left(\frac{C}{A}\right)^{2}{\bar v}-m\right]^{2}
\nonumber\\
& + &
\frac{4C^{2}}{A^{2}}\left(1+\frac{m}{2}\right)^{2}(\sin{\bar k_{n}})^{2}\bigg)^{1/2}\nonumber\\
\nonumber\\
&=&\prod_{{\vec x}}\prod_{\alpha=1}^{N_{c}}\left[\frac{2C}{\vert A
\vert}(1+m/2)\right]^{N_{\tau}}\times \bigg(2\cosh[N_{\tau}E]
\nonumber\\
& + & 2\cos[N_{\tau}(\theta_{a}({\vec x})-i\mu)]\bigg),
 \label{eqn25}
\end{eqnarray}
with
\begin{eqnarray}
E&=&{\rm arcsinh}
\vert\frac{2(1+m/2)\frac{d}{N_{c}}(\frac{C}{A})^{2}{\bar
v}-m}{\frac{2C}{A}(1+m/2)}\vert,
\nonumber\\
{\bar k_{n}}&=&k_{n}+\theta_{a}(x)-i\mu.
\label{eqn26}
\end{eqnarray}

We then perform the integration over $U_{4}$ and obtain
\begin{eqnarray}
&&\int D[U_{4}](\det\Delta)^{N_{f}/N_{\tau}}\nonumber\\
&=&\prod_{{\vec x}}\left[\frac{2C}{A}(1+m/2)\right]^{4N_{f}}
\nonumber\\
&\times& \left( 2\cosh(N_{c}\mu N_{\tau}) + \frac{\sinh
\left[(N_{c}+1)EN_{\tau}\right]}{\sinh
(EN_{\tau})}\right)^{N_{f}/N_{\tau}}.
\nonumber\\
 \label{eqn27}
\end{eqnarray}
As a result, the partition function (\ref{eqn17}) becomes
\begin{eqnarray}
Z&=&\exp \left(-\sum_{x}(1+m/2)^{2}\frac{dC^{2}}{N_{c}A^{2}}N_{f}({\bar v})^{2} \right)\nonumber\\
\nonumber\\
&\times&\prod_{{\vec x}}\left[\frac{2C}{A}(1+m/2)\right]^{4N_{f}}
\nonumber\\
&\times& \left( 2\cosh[N_{c}\mu
N_{\tau}]+\frac{\sinh[(N_{c}+1)EN_{\tau}]}{\sinh[EN_{\tau}]}\right)^{N_{f}/N_{\tau}}.
\nonumber\\
\label{eqn28}
\end{eqnarray}
Consequently, we obtain the effective free energy
\begin{eqnarray}
F_{eff}&=&
-\ln Z/(\sum_{\vec{x}})
\nonumber\\
&=&\left(1+\frac{m}{2}\right)^{2}\frac{dC^{2}}{N_{c}A^{2}}\left[N_{f}({\bar
v})^{2}\right]
\nonumber\\
&-&N_{f}T\times\ln \left(2\cosh \left({N_{c}\mu \over
T}\right)+\frac{\sinh \left( {(N_{c}+1)E \over T}\right)}{\sinh
\left({E \over T}\right)}\right)
\nonumber\\
&-&4N_{f}\ln\left(\frac{2C}{A}\left(1+\frac{m}{2}\right)\right),
\label{eqn29}
\end{eqnarray}
This formula is simple enough to make an analytical study of the
vacuum, helpful in understanding the phase structure of strong
coupling lattice QCD.

\section{Phase structure in the chiral limit}
\label{sec4}

\subsection{Chiral symmetry spontaneously breaking at $\mu=T=0$}

In case of $T=0$, the effective free energy (\ref{eqn29}) reduces
to a simpler form:
\begin{eqnarray}
\lim_{T\rightarrow 0}F_{eff}=\frac{dN_{f}}{N_{c}}({\bar v}^{'})^{2}-N_{c}N_{f}\max(\mu,E),
\label{eqn30}
\end{eqnarray}
with
\begin{eqnarray}
E={\rm arcsinh} \vert\frac{d}{N_{c}}{\bar v}^{'}\vert,
 \label{eqn31}
\end{eqnarray}
and
\begin{eqnarray}
{\bar v}^{'}=\vert\frac{C}{A}{\bar v}\vert.
 \label{eqn32}
\end{eqnarray}
At $\mu=0$, the effective free energy becomes
\begin{eqnarray}
F_{eff}[T=0,\mu=0]=\frac{dN_{f}}{N_{c}}({\bar v}^{'})^{2}-N_{c}N_{f}E. \label{eqn33}
\end{eqnarray}

The rescaled chiral condendate ${\bar v}^{'}$ is determined by the
conditions of minimizing the effective free energy
\begin{eqnarray}
\frac{\partial F_{eff}}{\partial{\bar
v}^{'}}=0,\frac{\partial^{2}F_{eff}}{\partial({\bar
v}^{'})^{2}}>0,
 \label{eqn34}
\end{eqnarray}
from which we get
\begin{eqnarray}
{\bar v}^{'}=\left(\frac{\sqrt{17}-1}{2}\right)^{1/2}=O(t^{0})\neq
0,
 \label{eqn35}
\end{eqnarray}
with the spatial dimension and number of colors taken to be $d=3$
and $N_{c}=3$ respectively. So the chiral condensate defined in
Eq. (\ref{eqn14}) is
\begin{eqnarray}
{\bar v}=\langle {\bar q}q \rangle =\langle {\bar
\psi}\left(1-\frac{a}{2}D\psi \right) \rangle
=O(t^{0})\vert\frac{A}{C}\vert\neq 0.
 \label{eqn36}
\end{eqnarray}
Therefore in the strong coupling limit, spontaneous symmetry
breaking of the extended chiral symmetry occurs at zero
temperature and zero chemical potential. This qualitatively agrees
with that obtained from the external source method\cite{Ichinose}.

\subsection{Chiral symmetry restoration at $\mu\neq 0$ and $T=0$}

For small $\mu$, the effective free energy is the same as Eq.
(\ref{eqn33}), i.e.
\begin{eqnarray}
F_{eff}=N_{f}{\bar v}'^{2}-3N_{f}E.
 \label{eqn38}
\end{eqnarray}
According to Eq. (\ref{eqn34}), the effective free energy reaches
its global minimum $F_{min}=-1.971N_{f}$ at ${\bar v}^{'}=1.25$.

For large enough $\mu$,  the effective free energy (\ref{eqn30})
reaches its minimum
\begin{eqnarray}
F_{min}=-3N_{f}\mu
 \label{eqn37}
\end{eqnarray}
when ${\bar v}^{'}=0$, i.e., the system is in the chiral symmetric
phase.

\begin{figure} [htbp]
\begin{center}
\includegraphics[totalheight=2.5in]{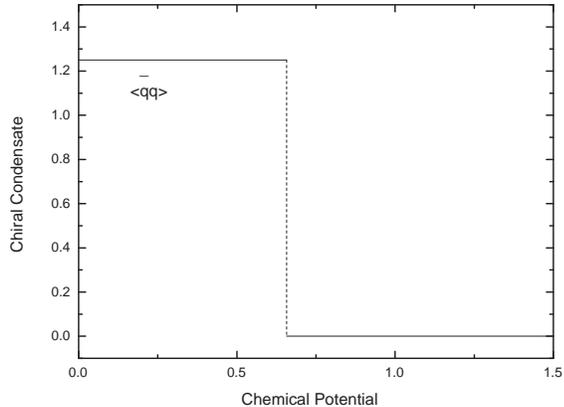}
\end{center}
\vspace{-1cm} \caption{Rescaled chiral condensate as a function of
$\mu$ at $T=0$. It drops discontinuously to zero at
$\mu=\mu_{c}=0.657$, which signals a first order phase
transition.} \label{fig1}
\end{figure}

Therefore, when $\mu$ increases from zero, the global minimum
changes from $-1.971N_{f}$ to $-3N_{f}\mu$ at the critical
chemical potential $\mu_c$, determined by
$-1.971N_{f}=-3N_{f}\mu_{c}$. Therefore $\mu_{c}=0.657$, at which
the order parameter ${\bar v}^{'}$ changes discontinuously, as
shown in Fig. \ref{fig1}. This is a clear indication of first
order chiral phase transition.

\subsection{Chiral symmetry restoration at $\mu=0$ and $T\neq 0$}

\begin{figure} [htbp]
\begin{center}
\includegraphics[totalheight=2.5in]{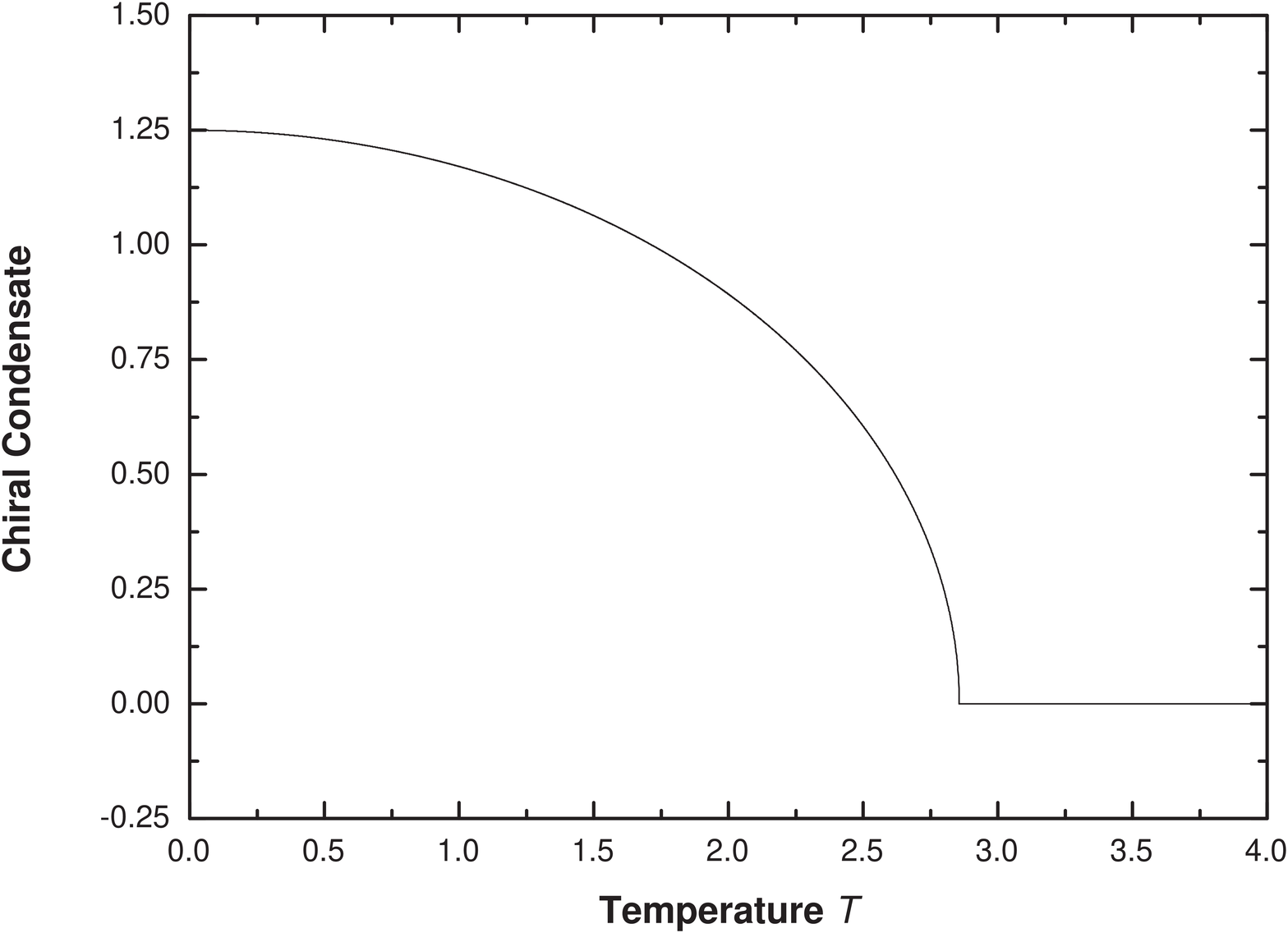}
\end{center}
\vspace{-1cm} \caption{Rescaled chiral condensate as a function of
$T$ at $\mu=0$. It decreases continuously to zero at $T=T_c$,
which signals a second order phase transition.} \label{fig2}
\end{figure}

In the small ${\bar v}^{'}$ region, the effective free energy
takes the form
\begin{eqnarray}
F_{eff}(\mu,T,{\bar v}^{'})&=&C_{0}(\mu,T)+C_{1}(\mu,T) {\bar
v}'^{2}
\nonumber \\
&+&C_{2}(\mu,T){\bar v}'^{4}+O({\bar v}'^{6}), \label{eqn39}
\end{eqnarray}
where
\begin{eqnarray}
C_{0}(\mu,T)&=&-N_{f}T \ln \left(2\cosh(3\mu/T)+4\right)
\nonumber\\
&-&4N_{f}\ln\left(\frac{2C}{A}\right),
\nonumber\\
C_{1}(\mu,T)&=&N_{f}\left(1-\frac{20}{5T+2T\cosh(3\mu/T)}\right),
\nonumber\\
C_{2}(\mu,T)&=&\frac{4N_{f}}{3T+2\cosh(3\mu/T)}
\nonumber\\
&\times&
\left(5+\frac{65-34\cosh(3\mu/T)}{T^{2}(5+2\cosh(3\mu/T)}\right).
 \label{eqn40}
\end{eqnarray}
According to the Landau's theory\cite{Landau}, a second order
phase transition occurs when the coefficient of ${\bar v}'^{2}$
becomes zero. In the case $\mu=0$ and $T\neq 0$, there is a
critical value for $T=T_c$ at which $C_{1}(0,T_{c})=0$. In this
case, ${\bar v}'$ vanishes continuously, implying  a second order
phase transition at the critical temperature $T_{c}$, as shown in
Fig. \ref{fig2}.

\subsection{Phase diagram on the ($\mu,T$) plane}

Finally we study the phase structure on the ($\mu,T$) plane, based
on above discussions. According to Eq. (\ref{eqn40}), the
coefficient of ${\bar v}'^{2}$ vanishes when
\begin{eqnarray}
\mu_{c}(T)=\frac{T_{c}}{3} {\rm arccosh}
\left(\frac{20-5T_{c}}{2T_{c}}\right), \label{eqn41}
\end{eqnarray}
which corresponds to the second order chiral phase transition
line. When the coefficient of ${\bar v}'^{4}$ becomes negative,
first order chiral phase transition occurs at some positive value
of $C_{1}$.

\begin{figure} [htbp]
\begin{center}
\includegraphics[totalheight=2.6in]{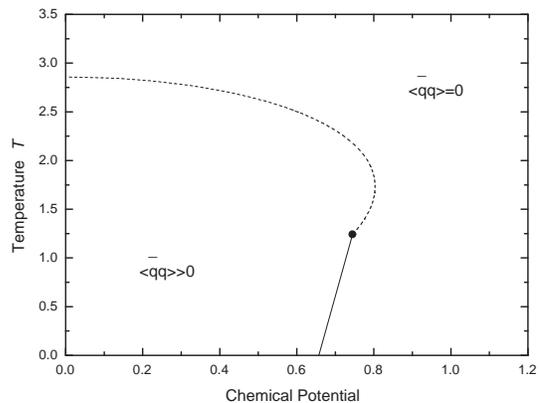}
\end{center}
\vspace{-1cm} \caption{Phase diagram on the $(\mu,T)$ plane. The
dotted and solid lines stand respectively for the second and first
order transitions. The circle is the tricritical point.}
\label{fig3}
\end{figure}

Consequently, we have a tricritical point
\begin{eqnarray}
T_{tri}=1.241, ~~~~~ \mu_{tri}=0.745.
 \label{eqn42}
\end{eqnarray}
when $C_{1}$ and $C_{2}$ vanish simultaneously. These results
indicate a second order chiral phase transition for $T\geq
T_{tri}$ and a first order one for $T\leq T_{tri}$, as shown in
Fig. \ref{fig3}. The existence of the tricritical point is
consistent with the recent result from the Hamiltonian approach
with Wilson fermion\cite{Luo:2004mc}.

\section{Discussion and outlook}
\label{sec5}

In the preceding sections, we have made the first attempt to
investigate the phase diagram  of Lagrangian lattice QCD with
overlap fermions at finite temperature $T$ and chemical potential
$\mu$. In the strong coupling limit, we discovered a tricritical
point on the ($\mu,T$) plane, separating the second order chiral
phase transition line at small $\mu$ and large $T$, and first
order chiral phase transition line at large $\mu$ and small $T$.

There are still some open questions. (1) The strong coupling is
far from the continuum limit. (2) As in Refs.
\cite{Ichinose,Nagao}, we considered only limited terms in Eq.
(\ref{eqn19}). One may have to include infinitive terms in order
to completely suppress the doublers. Analytical calculations will
be extremely difficult\cite{Ichinose,Nagao}, even at $\mu=T=0$.
Further study of QCD with overlap fermions at finite temperature
and chemical potential, ether analytical or numerical, will be
very interesting.


\acknowledgments

We would like to thank K. Nagao for useful discussions. This work
is supported by the Key Project of National Science Foundation
(10235040), Key Project of National Ministry of Eduction (105135),
Project of the Chinese Academy of Sciences (KJCX2-SW-N10) and
Guangdong Natural Science Foundation (05101821).

\end{document}